\documentclass[journal,web]{ieeecolor}
\usepackage{generic}
\usepackage{cite}
\usepackage{amsmath,amssymb,amsfonts}
\usepackage{algorithmic}
\usepackage{graphicx}
\usepackage{algorithm,algorithmic}
\usepackage{siunitx}
\usepackage{booktabs}
\usepackage{multirow}
\usepackage{makecell}
\usepackage{subcaption}
\usepackage{float}
\usepackage{hyperref}
\usepackage{tabularx}
\usepackage{array}
\usepackage{float}
\usepackage{booktabs}
\usepackage{tabularx}
\usepackage{array}

\hypersetup{hidelinks=true}
\usepackage{textcomp}
\def\BibTeX{{\rm B\kern-.05em{\sc i\kern-.025em b}\kern-.08em
    T\kern-.1667em\lower.7ex\hbox{E}\kern-.125emX}}
\begin{document}
\raggedbottom
\title{Human-Centered Explainable AI for TinyML Edge Devices: A
Pareto-Based Selection Framework with LLM-Guided Design}
\author{Zeinab Dehghani, Dhavalkumar Thakker, Koorosh Aslansefat, Kuniko Paxton, Bhupesh Kumar Mishra, Baseer Ahmad and Rameez Raja Kureshi
\thanks{Zeinab Dehghani, Dhavalkumar Thakker, Koorosh Aslansefat, Kuniko Paxton, Bhupesh Kumar Mishra, Baseer Ahmad, and Rameez Raja Kureshi are with the School of Digital and Physical Sciences, University of Hull, Cottingham Road, Hull, HU6 7RX, United Kingdom (e-mails: zeinab.dehghani068@gmail.com, D.Thakker@hull.ac.uk, e-mail: K.Aslansefat@hull.ac.uk, k.paxton@hull.ac.uk, Bhupesh.Mishra@hull.ac.uk, Baseer.Ahmad@hull.ac.uk, R.Kureshi@hull.ac.uk). Corresponding author: Kuniko Paxton
}}

\maketitle

\begin{abstract}
Edge Artificial Intelligence (Edge AI) enables the deployment of AI models directly on local edge devices, while such deployments are subject to strict resource constraints, particularly in clinical applications requiring local and timely inference. In such contexts, explainable artificial intelligence (XAI) can serve as a human-AI interface intended to support healthcare professionals' and patients' understanding of model predictions and informed decision-making. To fulfill this role, XAI method selection for TinyML deployments can be formulated as a human-centered multi-objective design problem that jointly considers qualitative stakeholder preferences, explanation quality, and proxy-based deployment cost. We propose a framework that integrates a large language model (LLM)-guided design interface that maps qualitative stakeholder preferences to candidate XAI methods, followed by deterministic feasibility filtering and Pareto-based optimization. The framework exposes trade-offs among explanation fidelity, stability, and proxy-based deployment cost while characterizing their implications for explanation quality and estimated deployment feasibility. A proof-of-concept evaluation on a skin lesion classification task illustrates how the framework systematically compares candidate XAI methods and identifies Pareto-efficient trade-offs. The present evaluation covers the computational selection stages, while physical MCU deployment and empirical human-expert validation remain outside the scope of this study.
\end{abstract}


\section{Introduction}
The use of artificial intelligence (AI) has expanded rapidly in healthcare. In particular, driven by the need to protect patient data privacy and the demand for real-time processing, there is growing interest in edge AI, which runs AI models on resource-constrained platforms such as microcontrollers (MCUs) and low-power embedded devices. In these environments, models must operate under strict latency, memory, and energy constraints, motivating active research in TinyML to maintain predictive performance under limited computational resources~\cite{alajlan2022tinyml,kallimani2024tinyml,neseem2024ai}.\\
On the other hand, medical AI requires not only predictive performance but also safety and reliability. Therefore, explainable artificial intelligence (XAI) is regarded as an important component for examining the basis of diagnostic predictions and identifying potential risks~\cite{rasheed2022explainable}. In fact, since clinicians and patients may use explanations to support their understanding of AI-assisted decisions, explanations can serve as an important interface that supports appropriate trust and informed use~\cite{bussone2015role,shin2021effects,rosenbacke2024explainable,chaddad2023survey}.\\
However, many existing XAI methods are designed with computationally rich environments in mind, and challenges remain regarding their application to TinyML environments. For example, activation- and gradient-based methods such as Class Activation Mapping (CAM)~\cite{zhou2016learning} and Gradient-weighted Class Activation Mapping (Grad-CAM)~\cite{selvaraju2017grad} require access to internal model information, while perturbation-based methods such as Local Interpretable Model-agnostic Explanations (LIME)~\cite{ribeiro2016should} and Randomized Input Sampling for Explanation (RISE)~\cite{petsiuk2018rise} involve multiple rounds of inference, resulting in high execution costs on devices~\cite{attioui2026sustainable,majdoubi2025avi}. In addition, hardware heterogeneity and environmental variability may affect explanation stability over time, reducing predictability in operational settings~\cite{vermesan2024explainability,neseem2024ai}. As a result, in real-world edge environments, the choice of explanation method often depends on the developer’s experience and judgment, and alignment with hardware constraints and operational requirements is not sufficiently considered.\\
Furthermore, XAI in healthcare is not merely a model analysis tool but can also function as a decision-support interface for healthcare professionals and patients. It is well known that the quality and presentation of explanations can influence user trust and reliance behavior, and inappropriate explanations can lead to over-reliance or erroneous judgments~\cite{bussone2015role,shin2021effects,leichtmann2023effects}. Nevertheless, existing research tends to treat the selection of explanation methods as a purely technical problem, and the systematic incorporation of intended-user needs and usage context into the selection process has received limited attention.\\
Moreover, with the diversification of XAI methods in recent years, selecting a method suited to the intended use and operational environment has become a critical challenge in itself. Recent research has highlighted the importance of human-centered XAI, emphasizing the selection and presentation of explanations that align with users' objectives and decision-making contexts~\cite{lai2023selective}. In other words, explanations should not only be technically correct but also understandable and usable by healthcare professionals and patients~\cite{spreeuwenberg2019choose}.\\
Thus, selecting an XAI method in a TinyML environment requires simultaneously considering diverse requirements, such as hardware constraints, the characteristics of the explanation method, explanation quality, task objectives, and human interpretability requirements. However, comprehensively evaluating these requirements and selecting an appropriate explanation method is not easy and places a significant burden even on experts. Recent advances in large language models (LLMs) have highlighted their potential to function as interfaces between human intent and system design~\cite{bommasani2021opportunities,openai2023gpt4}. In particular, LLMs can support the exploration, generation, and preliminary ranking of candidate solutions based on stakeholder requirements, thereby reducing the manual effort traditionally required to navigate complex design spaces while leaving optimization and feasibility enforcement to deterministic procedures~\cite{el2025can}.\\
Therefore, in this study, we formulate the selection of XAI methods in TinyML environments as a human-centered, constrained multi-objective optimization problem. The proposed framework evaluates trade-offs among explanation fidelity, stability, and proxy-based deployment cost using Pareto optimization, while leveraging LLMs to map qualitative stakeholder-oriented requirements, represented in this study through predefined goal profiles, to candidate XAI methods. The LLMs support the preliminary ranking of candidate methods according to qualitative stakeholder requirements, while deployment constraints are enforced separately through deterministic feasibility rules. Additionally, feasibility assessment and Pareto-based candidate identification are performed using deterministic procedures, supporting transparency and auditability. The computational selection stages of the framework are demonstrated through a proof-of-concept skin lesion classification study using the HAM10000 dataset~\cite{tschandl2018ham10000} and MobileNetV3-Small~\cite{howard2019searching}.

\section{Literature Review}
This section reviews prior work on (i) XAI deployment in TinyML environments, (ii) human-centered explainability, and (iii) frameworks for XAI method selection, including recent LLM-assisted approaches.

\subsection{Explainability Method Applications in TinyML}
This review also considers related work from the broader TinyML and edge AI domains because studies specifically addressing XAI in medical TinyML remain limited, while these broader studies address similar challenges in deploying explainability under resource constraints. Existing research has explored the use of XAI to support TinyML system design and optimization, including model pruning, hardware design, robustness enhancement, and anomaly detection~\cite{dehrouyeh2025pruning,dehrouyeh2025tinyml, lamaakal2026explainable,sabih2023robust,gulati2025nanoxai}. These studies primarily used XAI to support the design and optimization of TinyML systems rather than to provide explanations directly to end users. Consequently, the computational implications of user-facing XAI under on-device deployment have received limited attention. Other studies have integrated XAI into TinyML and edge AI applications, with explanations generated either on servers~\cite{arthi2024optimized,majdoubi2025enhancing} or directly on edge devices~\cite{patel2025xai,trivedi2025defeat,shah2025cosmolite}. However, studies in this category often provide limited analysis of the computational costs associated with XAI implementation and hardware constraints. Recent work has begun to consider hardware constraints and the computational cost of explanation generation in edge AI environments~\cite{uddin6225231edge,attioui2026sustainable,majdoubi2025avi}. Nevertheless, XAI selection in these studies remains largely developer-driven with limited consideration of stakeholder preferences or end-user evaluation of explanation quality. Furthermore, few studies explicitly incorporate stakeholder-oriented requirements into the design and selection of XAI methods for resource-constrained deployments.

\subsection{Human-Centered Explainability}
Explainability plays a critical role in shaping human interaction with AI systems, particularly in safety-critical domains such as healthcare~\cite{rasheed2022explainable}. Human-centered explainability focuses on how explanations are designed, presented, and interpreted by users within specific decision-making contexts~\cite{rong2023towards} and considers how users perceive and respond to explanations~\cite{kim2024human}. Empirical studies show that explanation quality can influence user trust and reliance, potentially leading to over-reliance, under-reliance, or inappropriate trust depending on the characteristics of the explanation~\cite{bussone2015role,shin2021effects}. However, trust is not synonymous with reliance, and this distinction underscores the necessity of employing appropriate and validated evaluation measures to guide the design and evaluation of human-centered AI systems~\cite{scharowski2023exploring}. These findings highlight that explanations should be evaluated not only in terms of technical correctness but also in terms of their potential effects on human judgment and decision-making. In medical applications, explanations should align with domain knowledge and support clinically meaningful reasoning~\cite{rosenbacke2024explainable}. However, existing work often treats human factors and system constraints separately, limiting the applicability of these approaches to real-world deployments.

\subsection{Selection of Explainability Methods}
Over the years, several frameworks have been proposed to support the selection of appropriate XAI techniques. Prior approaches include a multi-criteria decision-making framework for ranking candidate methods~\cite{matejova2025multi}, property-oriented selection frameworks targeting data scientists~\cite{vermeire2021choose}, conceptual frameworks for XAI centered on stakeholders and their explanatory needs~\cite{langer2021we,stodt2024demystifying}, and taxonomy-based approaches for organizing and comparing XAI techniques~\cite{zhukov2025explainable}. These studies have contributed to a more structured understanding and selection of XAI methods. That said, they typically rely on manual evaluation, focus primarily on developers or data scientists, or remain at the conceptual level with limited empirical involvement of domain experts. A recent study incorporated LLMs into the design of XAI for small MCUs~\cite{el2025can}. While this highlighted the potential of generative AI for automated design support, the resulting recommendations remain largely driven by hardware and model-performance considerations, with limited representation of domain-user requirements in the design process.\\
To address this gap, we propose a human-centered TinyML XAI selection framework that combines stakeholder-oriented goal profiles, LLM-guided candidate generation, deterministic feasibility assessment, Pareto-based method selection, and a proposed human-expert review stage. The present proof-of-concept study evaluates the computational selection stages, while the human-expert review stage remains to be empirically implemented and validated.

\subsection{Main Contributions}
In this manuscript, we propose four main contributions that support the selection of appropriate XAI configurations under task-specific and resource constraints. The framework is intended to enable transparent and resource-aware XAI design for resource-constrained edge AI applications. Its computational components are demonstrated through a proof-of-concept medical imaging case study.

\begin{enumerate}
    \item \textbf{Human-Centered, Stakeholder-Oriented Formulation:}
    We formulate XAI method selection as a constrained multi-objective optimization problem that jointly considers qualitative stakeholder priorities, explanation quality, and proxy-based deployment cost.
    
    \item \textbf{LLM-Guided Candidate Generation:}
    We introduce a mechanism that maps qualitative stakeholder preferences to candidate XAI methods, while feasibility assessment and Pareto-based evaluation remain governed by deterministic procedures.

    \item \textbf{Pareto-Based Selection:} We develop a Pareto-based selection framework that captures trade-offs among explanation fidelity, stability, and proxy-based deployment cost.
    
    \item \textbf{Transparent and Auditable Selection:}
    We provide a transparent and auditable comparison of Pareto-efficient XAI configurations to support subsequent human-expert review and deployment-oriented planning.
\end{enumerate}

\section{Proposed Method}
\begin{figure*}[t]
    \centering
    \includegraphics[width=\textwidth]{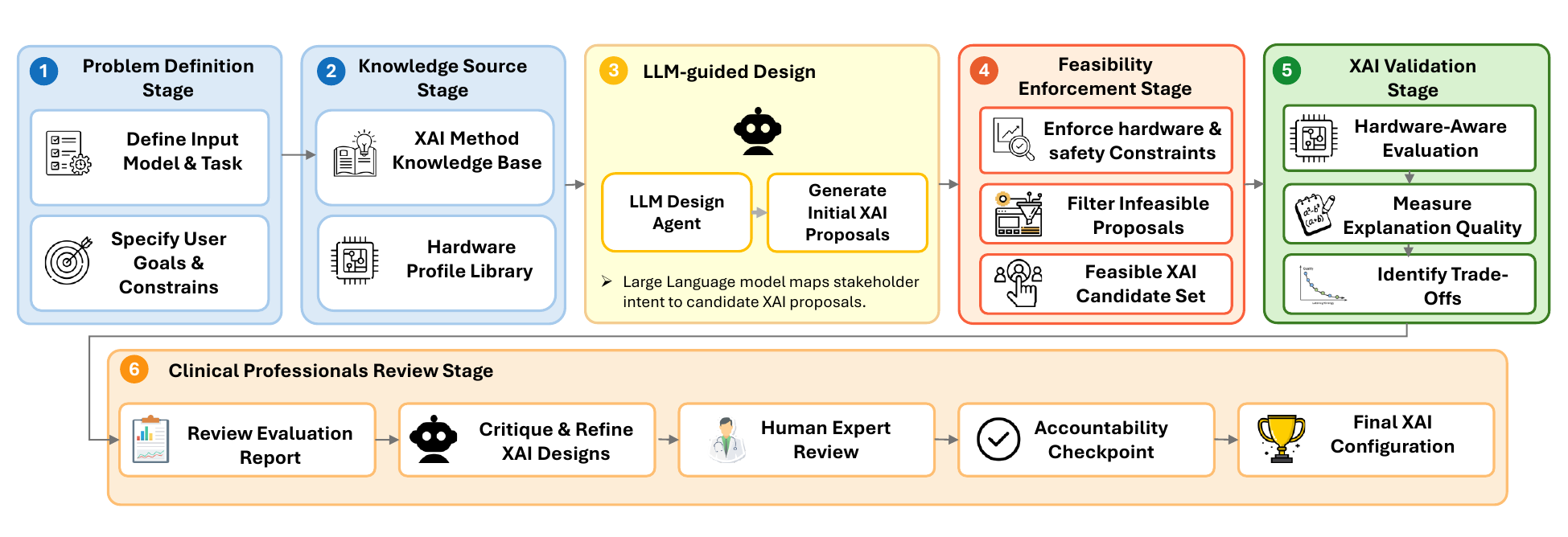}
    \caption{Overview of the proposed human-centered, hardware-profile-aware framework for selecting XAI methods for TinyML edge deployment scenarios. The framework integrates stakeholder-oriented goal profiles, LLM-guided candidate generation, deterministic feasibility enforcement, Pareto-based evaluation, and a proposed human-expert review stage for final method validation.}
    \label{fig:proposed_method}
\end{figure*}

This work proposes a human-centered, hardware-aware framework for selecting XAI methods for TinyML edge deployment scenarios, as illustrated in Fig.~\ref{fig:proposed_method}. 

\subsection{Problem Definition Stage}
The proposed framework begins with an explicit problem-definition stage in which the predictive-model specification and stakeholder intent are recorded independently of XAI method and hardware metadata. This separation is important for safety-sensitive XAI design because it prevents qualitative stakeholder requirements from being conflated with device-specific resource constraints~\cite{vermesan2024explainability,rasheed2022explainable}. The model specification describes the architecture family, task definition, input representation, and output format. Stakeholder intent is represented through three predefined goal profiles, Clinical, TinyML, and Balanced, which encode the decision context, intended users, primary and explanation goals, operational mode, time sensitivity, explanation-quality expectation, risk tolerance, explanation representation, and safety considerations. Deterministic rule-based mappings subsequently translate the profile fields into policy constraints used during candidate assessment and feasibility enforcement, following a structured approach suitable for human-centered and safety-critical AI systems~\cite{arrieta2020explainable,holzinger2018machine}.

\subsection{Knowledge Source Stage}

The construction of candidate explanation pipelines is supported by two
structured knowledge sources: an XAI method knowledge base and a hardware
profile library.

\subsubsection{XAI Method Knowledge Base}

The XAI method knowledge base records the properties required for
candidate enrichment and deterministic feasibility assessment. These
properties include method family, execution scope, required access to
model weights, activations, or gradients, the number of forward passes,
and available analytical resource proxies. The resource fields are used
for relative comparison and feasibility screening rather than as direct
measurements of runtime, energy consumption, or physical memory usage.
The knowledge base also distinguishes methods intended for on-device
execution from host-only methods and records the requirements associated
with each method.

\subsubsection{Hardware Profile Library}
\label{sec:hardware_library}

The hardware profile library represents the target execution environment,
including supported numerical precision, runtime backend, SRAM and Flash
capacity, system-memory reservations, stack and heap reservations, and
model-memory reservations. Given an instantiated target profile, the
remaining SRAM and Flash headroom available for XAI is calculated
deterministically.

\subsection{LLM-Guided Design Stage}

The LLM receives the predictive-model specification, the qualitative
stakeholder goal profile, and a structured summary of the complete XAI
method catalog. The catalog summary includes method family, execution
scope, and method requirements, but does not provide measured latency,
energy, SRAM, or Flash values. The LLM is instructed to return a ranked
shortlist containing no more than five methods and to align the ranking
with the qualitative stakeholder profile.

The prompt prevents the LLM from generating resource estimates or enforcing
hardware feasibility. Method metadata and deployment
constraints are attached and evaluated deterministically in the subsequent
stage. A separate Pareto-constrained LLM call is used during final
configuration selection, as described in
Section~\ref{sec:final_selection}.

\subsection{Feasibility Enforcement Stage}

Let $M=\{m_1,\ldots,m_N\}$ denote the ranked methods proposed by the LLM.
Each candidate is first enriched using the method knowledge base, which
supplies its execution scope, model-access requirements, forward-pass
requirement, and available resource proxies.

Deterministic filtering is then performed in two stages. The first stage
checks SRAM and Flash limits whenever both the candidate estimate and the
corresponding hardware limit are defined. The second stage applies
execution-scope and stakeholder-derived policy constraints, including
forward-pass limits and the permitted use of multi-pass methods.

A host-only method is never classified as MCU-feasible. When host fallback
is permitted, a host-only method may be retained as a hybrid alternative,
but it is not included in the MCU-feasible set. When host fallback is not
permitted, the method is rejected from the intended deployment.

Let $\mathcal{C}_{\mathrm{MCU}}$ denote the set of applicable MCU
constraints. The MCU-feasibility indicator is defined in Eq.~\ref{eq:mcu_feasible}.

\begin{equation}
\label{eq:mcu_feasible}
\mathbb{I}_{\mathrm{MCU}}(m_i)
=
\begin{cases}
1, & \text{if }m_i\text{ satisfies every applicable }
c\in\mathcal{C}_{\mathrm{MCU}},\\
0, & \text{otherwise}.
\end{cases}
\end{equation}

The MCU-feasible method set is therefore defined in Eq.~\ref{eq:mcu_set}.

\begin{equation}
\label{eq:mcu_set}
M_{\mathrm{MCU}}
=
\left\{
m_i\in M
\mid
\mathbb{I}_{\mathrm{MCU}}(m_i)=1
\right\}.
\end{equation}

Fidelity and stability are evaluated across the complete parameterized
configuration set. For each stakeholder profile, however, only
configurations whose base method belongs to $M_{\mathrm{MCU}}$ and for
which all required objective values are available are included in the
profile-specific deployment-cost and Pareto analyses.

\subsection{XAI Validation Stage}

The validation stage evaluates explanation fidelity, explanation
stability, and relative deployment cost. These quantities are subsequently
used to identify non-dominated trade-offs among the feasible
configurations.

\subsubsection{Measure Explanation Quality}

Explanation quality is evaluated from two complementary perspectives:
attribution fidelity and explanation stability. Fidelity measures the
extent to which highly attributed image regions affect the model score.
Stability measures the structural consistency of the resulting heatmaps
under mild input variations
\cite{velmurugan2021developing,miro2024assessing,
alvarez2018robustness}.

\textbf{Fidelity Score:}

Because the case study addresses skin-lesion image classification using
heatmap-based explanations, attribution fidelity is evaluated using
deletion AUC, insertion AUC, and Area Over the Perturbation Curve (AOPC)
\cite{petsiuk2018rise,samek2016evaluating}.

For each input image
$x\in\mathbb{R}^{H\times W\times3}$, the evaluation class is fixed to the
class predicted from the unmodified input, as defined in Eq.~\ref{eq:predicted_class}:

\begin{equation}
\label{eq:predicted_class}
c^\star(x)=\arg\max_k z_k(x),
\end{equation}

where $z_k(x)$ denotes the logit of class $k$. The fixed-class score used
throughout the perturbation process is defined in Eq.~\ref{eq:fixed_class_score}.

\begin{equation}
\label{eq:fixed_class_score}
S_x(x')=z_{c^\star(x)}(x'),
\end{equation}

where $x'$ denotes the original image or one of its perturbed versions.
The implementation uses logits rather than softmax probabilities.

Let $h_m(x)$ denote the heatmap produced by configuration $m$. The
heatmap is resized to the input resolution. Its absolute values are then
independently min--max normalized for each input, as shown in Eq.~\ref{eq:heatmap_normalization}:

\begin{equation}
\label{eq:heatmap_normalization}
\bar{h}_m(x)
=
\frac{
\left|h_m(x)\right|
-
\min\left|h_m(x)\right|
}{
\max\left|h_m(x)\right|
-
\min\left|h_m(x)\right|
+
\epsilon
}.
\end{equation}

The normalized heatmap is partitioned into non-overlapping patches of
size $p\times p$. In the present implementation, $p=16$. For the
$224\times224$ inputs, the number of patches is calculated using Eq.~\ref{eq:number_of_patches}:

\begin{equation}
\label{eq:number_of_patches}
P
=
\frac{224}{16}
\times
\frac{224}{16}
=
196
\end{equation}

patches. The patches are ranked in descending order according to their
mean absolute attribution magnitude.

The evaluation uses $L=20$ perturbation intervals rather than applying
one perturbation step for every patch. At interval
$\ell\in\{0,\ldots,L\}$, the cumulative number and fraction of modified patches are defined in Eq.~\ref{eq:perturbation_intervals}.

\begin{equation}
\label{eq:perturbation_intervals}
q_\ell
=
\operatorname{round}
\left(
\frac{\ell P}{L}
\right),
\qquad
u_\ell
=
\frac{q_\ell}{P}.
\end{equation}

Let $\pi_{q_\ell}(x,h_m)$ denote the set containing the
$q_\ell$ highest-ranked patches. A blurred version of the input is used
as the baseline image $x_{\mathrm{base}}$. The implementation constructs
this baseline using a $21\times21$ uniform depthwise blur.

During deletion, the highest-ranked patches are progressively replaced
by the corresponding regions of the baseline image, as shown in Eq.~\ref{eq:deletion}:

\begin{equation}
\label{eq:deletion}
x^{(\ell)}_{\mathrm{del}}
=
\mathrm{Replace}
\left(
x,
x_{\mathrm{base}},
\pi_{q_\ell}(x,h_m)
\right).
\end{equation}

Insertion begins with the blurred baseline and progressively restores the
highest-ranked regions from the original image, as shown in Eq.~\ref{eq:insertion}:

\begin{equation}
\label{eq:insertion}
x^{(\ell)}_{\mathrm{ins}}
=
\mathrm{Replace}
\left(
x_{\mathrm{base}},
x,
\pi_{q_\ell}(x,h_m)
\right).
\end{equation}

The deletion and insertion curves are defined in Eq.~\ref{eq:curves}.

\begin{equation}
\label{eq:curves}
D_m(u_\ell)
=
S_x
\left(
x^{(\ell)}_{\mathrm{del}}
\right),
\qquad
I_m(u_\ell)
=
S_x
\left(
x^{(\ell)}_{\mathrm{ins}}
\right).
\end{equation}

The areas under these curves are numerically calculated using the
trapezoidal rule, as shown in Eqs.~\ref{eq:deletion_auc} and~\ref{eq:insertion_auc}:

\begin{equation}
\label{eq:deletion_auc}
\mathrm{AUC}_{\mathrm{del}}(m)
=
\sum_{\ell=1}^{L}
\frac{
D_m(u_{\ell-1})+D_m(u_\ell)
}{2}
\left(
u_\ell-u_{\ell-1}
\right),
\end{equation}

\begin{equation}
\label{eq:insertion_auc}
\mathrm{AUC}_{\mathrm{ins}}(m)
=
\sum_{\ell=1}^{L}
\frac{
I_m(u_{\ell-1})+I_m(u_\ell)
}{2}
\left(
u_\ell-u_{\ell-1}
\right).
\end{equation}

Lower deletion AUC and higher insertion AUC indicate better attribution
fidelity because removing highly attributed regions should rapidly reduce
the fixed-class score, whereas inserting them should rapidly restore it
\cite{petsiuk2018rise,hooker2019benchmark}.

AOPC measures the average reduction in the original fixed-class logit
across the $L$ non-zero deletion intervals, as defined in Eq.~\ref{eq:aopc}:

\begin{equation}
\label{eq:aopc}
\mathrm{AOPC}(m)
=
\frac{1}{L}
\sum_{\ell=1}^{L}
\left[
S_x(x)
-
D_m(u_\ell)
\right].
\end{equation}

Higher AOPC indicates that removing highly ranked regions causes a larger
reduction in the score of the original predicted class
\cite{samek2016evaluating}.

The image-level deletion AUC, insertion AUC, and AOPC values are averaged
over the evaluation dataset for every parameterized configuration. Since
the three metrics have different numerical ranges and optimization
directions, their configuration-level means are independently min--max
normalized across the evaluated configuration set. The normalized
quantities are denoted in Eq.~\ref{eq:normalized_fidelity_metrics}.

\begin{equation}
\label{eq:normalized_fidelity_metrics}
\widetilde{\mathrm{AUC}}_{\mathrm{del}}(m),
\qquad
\widetilde{\mathrm{AUC}}_{\mathrm{ins}}(m),
\qquad
\widetilde{\mathrm{AOPC}}(m).
\end{equation}

Because deletion AUC is minimized, its normalized value is inverted. The
composite fidelity score is then calculated using equal weights, as shown in Eq.~\ref{eq:fidelity_score}:

\begin{equation}
\label{eq:fidelity_score}
F(m)
=
\frac{1}{3}
\left[
\widetilde{\mathrm{AUC}}_{\mathrm{ins}}(m)
+
\widetilde{\mathrm{AOPC}}(m)
+
1-
\widetilde{\mathrm{AUC}}_{\mathrm{del}}(m)
\right].
\end{equation}

Higher values of $F(m)$ indicate better composite attribution fidelity
across the deletion, insertion, and AOPC criteria.

\textbf{Stability:}

Explanation stability is evaluated using three mild perturbed versions of
each input image. Each perturbation applies reflection padding of six
pixels followed by a random crop back to the original input dimensions.
This introduces a small spatial translation. Zero-mean Gaussian noise
with a standard deviation of $0.02$ is then added in the model-input
value scale.

The implementation does not explicitly verify preservation of the
predicted class after perturbation. These transformations are therefore
treated as mild input variations rather than guaranteed label-preserving
perturbations. Brightness and contrast transformations are not used in
the stability calculation.

For each original and perturbed input, the explanation heatmap is resized
to the input resolution and independently min--max normalized to
$[0,1]$. Let $h_m(x)$ denote the resulting heatmap for the original
input and let $h_m(\tilde{x}_j)$ denote the heatmap for the $j$th
perturbed input.

Structural similarity is measured using the Structural Similarity Index
Measure (SSIM) \cite{wang2004image,peng2020implementation}. For two
heatmaps $a$ and $b$, SSIM is defined in Eq.~\ref{eq:ssim}.

\begin{equation}
\label{eq:ssim}
\mathrm{SSIM}(a,b)
=
\frac{
(2\mu_a\mu_b+C_1)
(2\sigma_{ab}+C_2)
}{
(\mu_a^2+\mu_b^2+C_1)
(\sigma_a^2+\sigma_b^2+C_2)
},
\end{equation}

where $\mu_a$ and $\mu_b$ denote local means,
$\sigma_a^2$ and $\sigma_b^2$ denote local variances,
$\sigma_{ab}$ denotes local covariance, and $C_1$ and $C_2$ are
stabilizing constants.

In the present implementation, $J=3$ perturbed inputs are generated for
each image. The image-level stability score is defined in Eq.~\ref{eq:ssim_stability}.

\begin{equation}
\label{eq:ssim_stability}
\mathrm{Stab}_m(x)
=
\frac{1}{J}
\sum_{j=1}^{J}
\mathrm{SSIM}
\left(
h_m(x),
h_m(\tilde{x}_j)
\right).
\end{equation}

The configuration-level stability objective is obtained by averaging over
the complete evaluation dataset $\mathcal{D}$, as shown in Eq.~\ref{eq:configuration_stability}:

\begin{equation}
\label{eq:configuration_stability}
S(m)
=
\frac{1}{|\mathcal{D}|}
\sum_{x\in\mathcal{D}}
\mathrm{Stab}_m(x).
\end{equation}

Higher values of $S(m)$ indicate greater structural consistency of the
explanations under the applied input variations. Stability is considered
jointly with fidelity because strongly compressed or nearly invariant but
uninformative heatmaps may obtain high SSIM values.

\subsubsection{Identification of Trade-Offs}

Deployment overhead is represented using runtime and SRAM proxies rather
than direct measurements of physical energy consumption. For the
CAM-family configurations included in the profile-specific Pareto
analysis, the runtime proxy is defined in Eq.~\ref{eq:time_proxy}.

\begin{equation}
\label{eq:time_proxy}
T(m)
=
\overline{t}_{\mathrm{shared\_fwd}}
+
\overline{t}_{\mathrm{CAM}}
+
\overline{t}_{\mathrm{post}}(m),
\end{equation}

where $\overline{t}_{\mathrm{shared\_fwd}}$ is the mean time of the
shared model forward pass,
$\overline{t}_{\mathrm{CAM}}$ is the mean time required to construct the
base CAM from the shared activations and logits, and
$\overline{t}_{\mathrm{post}}(m)$ is the mean configuration-specific
post-processing time. For the unmodified CAM configuration,
$\overline{t}_{\mathrm{post}}(m)=0$.

An SRAM proxy, denoted by $M_{\mathrm{tensor}}(m)$, is obtained from the
analytical resource metadata in the XAI method knowledge base. The proxy
represents the estimated tensor storage associated with the base method.
In the current implementation, parameterized configurations belonging to
the same base method may share the same SRAM proxy. It should therefore
not be interpreted as a parameter-specific measured memory footprint.

Within each stakeholder profile, the runtime proxy is min--max normalized
over configurations with available timing values, whereas the SRAM proxy
is min--max normalized over configurations with available analytical SRAM
metadata in the merged profile table. Pareto filtering is subsequently
restricted to configurations with complete fidelity, stability, runtime,
and SRAM-derived deployment-cost values.

The relative deployment-cost proxy is defined in Eq.~\ref{eq:deployment_cost_proxy}.

\begin{equation}
\label{eq:deployment_cost_proxy}
C_g(m)
=
w_T\,
\mathrm{mm}_g
\left(
T(m)
\right)
+
w_M\,
\mathrm{mm}_g
\left(
M_{\mathrm{tensor}}(m)
\right).
\end{equation}

The weights are fixed as shown in Eq.~\ref{eq:deployment_cost_weights}.

\begin{equation}
\label{eq:deployment_cost_weights}
w_T=0.85,
\qquad
w_M=0.15,
\end{equation}

to prioritize the runtime component while retaining a contribution from
the SRAM proxy. Lower values of $C_g(m)$ indicate lower estimated
deployment cost. This quantity is a relative proxy and is not a
measurement of physical energy consumption on an MCU.

The three optimization objectives are therefore to maximize composite
fidelity $F(m)$, maximize stability $S(m)$, and minimize deployment cost
$C_g(m)$. For two feasible configurations $m_i$ and $m_j$ in profile
$g$, configuration $m_i$ dominates configuration $m_j$ according to Eq.~\ref{eq:dominance}.

\begin{equation}
\label{eq:dominance}
F(m_i)\geq F(m_j),
\qquad
S(m_i)\geq S(m_j),
\qquad
C_g(m_i)\leq C_g(m_j),
\end{equation}

with at least one strict inequality. The profile-specific Pareto set is defined in Eq.~\ref{eq:pareto_set}.

\begin{equation}
\label{eq:pareto_set}
\mathcal{P}_g
=
\left\{
m\in\mathcal{V}_g
\mid
\nexists\,m'\in\mathcal{V}_g
\text{ such that }m'\text{ dominates }m
\right\},
\end{equation}

where $\mathcal{V}_g$ denotes the valid feasible configuration set for
profile $g$. Every configuration in $\mathcal{P}_g$ represents a
non-dominated trade-off among fidelity, stability, and relative
deployment cost \cite{deb2011multi}.

\subsection{Pareto-Constrained Final Selection}
\label{sec:final_selection}

The final decision stage operates exclusively on the profile-specific
three-objective Pareto set $\mathcal{P}_g$. Configurations excluded by
feasibility enforcement or dominated during Pareto filtering are not
permitted to re-enter the selection process.

Within each Pareto set, fidelity, stability, and deployment cost are
independently min--max normalized. Let
$\widehat{F}_g(m)$, $\widehat{S}_g(m)$, and $\widehat{C}_g(m)$ denote the
normalized values for configuration $m\in\mathcal{P}_g$. The
profile-specific goal score is defined as

\begin{equation}
\label{eq:goal_score}
\begin{aligned}
R_g(m)
&=
\alpha_g\widehat{F}_g(m)
+
\beta_g\widehat{S}_g(m)
\\
&\quad+
\gamma_g
\left(
1-\widehat{C}_g(m)
\right),
\qquad
m\in\mathcal{P}_g.
\end{aligned}
\end{equation}

The profile-specific weights are

\begin{equation}
\label{eq:goal_weights}
(\alpha_g,\beta_g,\gamma_g)
=
\begin{cases}
(0.45,\,0.40,\,0.15),
& g=\mathrm{Clinical},\\
(0.20,\,0.20,\,0.60),
& g=\mathrm{TinyML},\\
(0.34,\,0.33,\,0.33),
& g=\mathrm{Balanced}.
\end{cases}
\end{equation}

The Clinical profile places the greatest emphasis on fidelity and
stability, the TinyML profile prioritizes deployment efficiency, and the
Balanced profile assigns approximately equal importance to the three
objectives. 

The ranked Pareto configurations, together with their fidelity, stability,
deployment-cost, and goal-score values, are supplied to GPT-4.1 mini with a
temperature of $0.1$. The API request uses the
\texttt{gpt-4.1-mini} model identifier, with the executed responses
recorded as \texttt{gpt-4.1-mini-2025-04-14}. The structured-output schema
restricts every method field to exact configuration identifiers from
$\mathcal{P}_g$.

For the Clinical profile, the LLM must select exactly one primary
configuration and no fallback. The selected configuration must have the
highest fidelity in $\mathcal{P}_g$; fidelity ties are resolved by lower
deployment cost and then higher stability.

For the Balanced and TinyML profiles, the LLM must select two distinct
primary configurations and one distinct fallback. One primary
configuration must be the maximum-fidelity, lowest-cost anchor. The second,
complementary primary configuration must satisfy

\begin{equation}
\label{eq:complementary_constraints}
\begin{aligned}
F(m_{\mathrm{comp}})
&\geq 0.70, \\
S(m_{\mathrm{comp}})
&\geq S(m_{\mathrm{anchor}})+0.02, \\
C(m_{\mathrm{comp}})
&\leq C(m_{\mathrm{anchor}})+0.05.
\end{aligned}
\end{equation}

The fallback must differ from both primary configurations, provide higher
fidelity than the complementary primary configuration, and remain within
the same low-cost interval:

\begin{equation}
\label{eq:fallback_constraints}
\begin{aligned}
F(m_{\mathrm{fallback}})
&> F(m_{\mathrm{comp}}), \\
C(m_{\mathrm{fallback}})
&\leq C(m_{\mathrm{anchor}})+0.05.
\end{aligned}
\end{equation}

The structured LLM response is subjected to deterministic validation. The
validator checks the stakeholder profile, exact Pareto membership,
configuration cardinality, duplication, anchor selection, and all numerical
constraints in Eqs.~\ref{eq:complementary_constraints} and
\ref{eq:fallback_constraints}. A response that fails validation is returned
to the LLM together with the detected errors for correction. 
A response that fails validation is returned to the LLM together with the detected errors for correction. If a valid response is not obtained, the procedure terminates without reporting a final selection.

\subsection{Proposed Human Review and Method Finalization}

The framework includes a proposed human-expert review stage to ensure that
final explanation selection is not based solely on quantitative
optimization. In a future clinical deployment, the Pareto-valid
explanations would be reviewed by medical experts using clinically
relevant criteria, including consistency with established dermatological
patterns, localization of salient regions to medically relevant areas,
and the absence of spurious artifacts or background bias.

The review stage is intended to identify explanations that may achieve
strong quantitative scores while remaining clinically misleading or
difficult to interpret. Human-centred evaluation can influence trust,
usability, and decision quality in AI-assisted systems
\cite{bussone2015role,leichtmann2023effects,
holzinger2018machine,doshi2017towards}.

This human-expert review stage was not empirically implemented in the
present proof-of-concept study and remains part of future validation.

\section{Experimental Setup}

The experimental study focuses on skin-lesion classification using three
stakeholder profiles: (1) Clinical, which prioritizes explanation fidelity
and stability; (2) TinyML, which prioritizes on-device deployability; and
(3) Balanced, which represents a compromise among explanation quality and
deployment cost. Because the predictive task is image classification, the
required explanation representation is a class-specific spatial heatmap.

The LLM proposal stage considers the complete structured XAI method catalog.
A separate parameterized configuration space is then constructed for the
quantitative fidelity and stability evaluation. For the profile-specific
timing and Pareto analyses, this configuration space is filtered according
to the base methods retained by the deterministic feasibility stage.

\subsection{Data, Model and Training Methods}

The proposed framework is evaluated on the HAM10000 skin-lesion dataset
\cite{tschandl2018ham10000,tschandl2020human}. The dataset contains
10,015 dermoscopic images belonging to seven diagnostic categories.
MobileNetV3-Small \cite{howard2019searching} is used as the predictive
backbone because it is designed for computationally constrained
environments.

The images are divided using stratified sampling with a fixed random seed
of 42. The resulting subsets contain 7,010 training images, 1,001
validation images, and 2,004 test images, corresponding approximately to
70\%, 10\%, and 20\% of the dataset, respectively.

\begin{table}[H]
\centering
\caption{Summary of the experimental setup.}
\label{tab:exp_setup}
\footnotesize
\setlength{\tabcolsep}{3pt}
\renewcommand{\arraystretch}{1.08}

\begin{tabularx}{\columnwidth}{
    >{\raggedright\arraybackslash}p{0.30\columnwidth}
    >{\raggedright\arraybackslash}X
}
\toprule
\textbf{Component} & \textbf{Description} \\
\midrule

Dataset
& HAM10000; 10,015 images and seven classes \\

Split
& Stratified training/validation/test split:
7,010/1,001/2,004 images ($70\%/10\%/20\%$) \\

Random seed
& 42 \\

Input
& $224\times224$ RGB images; MobileNetV3 preprocessing \\

Batch size
& 64 for model training and evaluation \\

Backbone
& MobileNetV3-Small with ImageNet-pretrained weights \\

Training stage 1
& Frozen backbone; two epochs; Adam with
learning rate $10^{-3}$ \\

Training stage 2
& Last approximately 40\% of backbone layers unfrozen;
two epochs; Adam with learning rate $3\times10^{-5}$ \\

Loss and imbalance handling
& Categorical cross-entropy with label smoothing of 0.05
and balanced class weights \\

Augmentation
& Horizontal flip; rotation factor 0.10; zoom factor 0.20;
contrast factor 0.20 \\

Evaluated XAI families
& CAM, Tiny-Saliency, LR-CAM, Micro-CAM, Binary-CAM,
TopK-CAM, and TopK$\times$Binary \\

Full XAI configuration set
& 67 parameterized configurations \\

Execution model
& One shared model forward pass followed by
configuration-specific post-processing \\

LLMs
& GPT-4.1 mini and Gemini 2.0 Flash; Gemini evaluated
in a separate run \\

GPT configuration
& OpenAI API; temperature 0.1 \\

Hardware profile
& Generic Cortex-M7 profile with 512\,kB SRAM and
2\,MB Flash \\

Explanation metrics
& Deletion AUC, insertion AUC, AOPC, and SSIM \\

Optimization objectives
& Composite fidelity, SSIM-based stability, and relative
deployment-cost proxy \\

\bottomrule
\end{tabularx}
\end{table}
All images are resized to $224\times224$ pixels and processed using the
MobileNetV3 preprocessing function. The training set is augmented using
horizontal flipping, a rotation factor of 0.10, a zoom factor of 0.20,
and a contrast factor of 0.20. Augmentation is not applied to the
validation and test sets.

A two-stage transfer-learning procedure is used. During the first stage,
the ImageNet-pretrained MobileNetV3-Small backbone is frozen and the
classification head is trained for two epochs using Adam with a learning
rate of $10^{-3}$. During the second stage, approximately the last 40\%
of the backbone layers are unfrozen and the model is fine-tuned for two
additional epochs using a learning rate of $3\times10^{-5}$. Both stages
use class weighting and categorical cross-entropy with label smoothing
of 0.05. A \texttt{ReduceLROnPlateau} callback monitors validation loss.
The principal experimental settings are summarized in
Table~\ref{tab:exp_setup}.
\subsection{Explainability Methods}

All 67 explanation configurations are implemented using a shared-forward
execution model. A single model inference returns the final convolutional
activation tensor and the class logits. The base CAM is calculated from
these shared outputs, after which the parameterized explanation variants
are generated using configuration-specific post-processing. Tiny-Saliency
is also generated from the shared activation tensor and does not require
an additional model inference.

The complete set of 67 configurations is evaluated for fidelity and
stability under each stakeholder profile. Before the profile-specific
deployment-cost and Pareto analyses, the configuration set is filtered
according to the base methods retained by the deterministic feasibility
stage. The complete parameterized design space is reported in
Table~\ref{tab:xai-configurations}.

\begin{table}[H]
\centering
\caption{Parameterized XAI configuration space. The symbols $d$, $s$,
$\tau$, and $k$ denote the LR-CAM downsampling factor, requested
Micro-CAM spatial size, Binary-CAM threshold, and retained TopK-CAM
activation ratio, respectively. Requested Micro-CAM sizes greater than
the $7\times7$ feature-map resolution are clipped to seven by the
implementation.}
\label{tab:xai-configurations}
\footnotesize
\setlength{\tabcolsep}{2.5pt}
\renewcommand{\arraystretch}{1.08}

\begin{tabular}{p{1.65cm} p{2.55cm} p{2.45cm} c}
\toprule
\textbf{XAI family}
& \textbf{Parameter}
& \textbf{Values}
& \textbf{No.} \\
\midrule

CAM
& --
& Baseline
& 1 \\

Tiny-Saliency
& --
& Baseline
& 1 \\

LR-CAM
& Downsampling $d$
& $1,\ldots,10$
& 10 \\

Micro-CAM
& Requested size $s$
& $1,\ldots,10$
& 10 \\

Binary-CAM
& Threshold $\tau$
& $0.1,\ldots,0.9$
& 9 \\

TopK-CAM
& Retained ratio $k$
& $0.02,0.04,\ldots,0.50$
& 25 \\

TopK$\times$Binary
& $k$ and $\tau$
& 11 parameter pairs
& 11 \\

\midrule
\textbf{Total}
& --
& --
& \textbf{67} \\

\bottomrule
\end{tabular}
\end{table}

For TopK$\times$Binary, the parameter grid is constructed from

\begin{equation}
k\in\{0.05,0.15,0.30,0.50\},
\qquad
\tau\in\{0.3,0.5,0.7\}.
\end{equation}

Eleven of the twelve Cartesian-product combinations are evaluated; the
combination $(k,\tau)=(0.50,0.7)$ is not included. For every included
combination, TopK-CAM is first applied to the base CAM, followed by the
binary-threshold operation.

LR-CAM reduces the $7\times7$ CAM resolution according to

\begin{equation}
H_d
=
\max
\left(
1,
\left\lfloor\frac{7}{d}\right\rfloor
\right),
\qquad
W_d
=
\max
\left(
1,
\left\lfloor\frac{7}{d}\right\rfloor
\right).
\end{equation}

Micro-CAM resizes the base CAM to

\begin{equation}
s_{\mathrm{eff}}
=
\min(s,7),
\end{equation}

which means that requested sizes $s=7,\ldots,10$ produce the same
effective $7\times7$ spatial resolution. Binary-CAM thresholds the
normalized base CAM at $\tau$. TopK-CAM retains at least one activation
and otherwise retains

\begin{equation}
\max
\left(
1,
\left\lfloor49k\right\rfloor
\right)
\end{equation}

of the 49 CAM locations.

\subsection{LLM-Guided Models}

The explanation-proposal stage uses OpenAI's GPT-4.1 mini and Gemini 2.0
Flash \cite{comanici2025gemini}. GPT-4.1 mini is accessed through the
OpenAI API using a temperature of 0.1. Gemini 2.0 Flash is evaluated in a
separate run using the same prompt structure.

For each stakeholder profile, the prompt contains the predictive-model
specification, the qualitative stakeholder goal profile, and a structured
summary of the XAI method catalog. The method summary includes method
family, execution scope, and method requirements such as the number of
forward passes. Explicit SRAM, Flash, latency, and energy values are not
provided to the LLM. The prompt instructs the LLM to propose and rank no
more than five methods without performing hardware-feasibility filtering
or generating resource estimates.

\subsection{Hardware and Deployment Constraints}

The proof-of-concept uses an instantiated generic Cortex-M7 hardware
profile with 512\,kB of total SRAM and 2,048\,kB of total Flash. The
profile reserves 64\,kB of SRAM for the system, 64\,kB for the stack and
heap, and predefined model footprints of 200\,kB SRAM and 1,500\,kB
Flash. Under these predefined values, the resulting headroom available
for XAI is

\begin{equation}
512-64-64-200
=
184\,\mathrm{kB}
\end{equation}

of SRAM and

\begin{equation}
2048-1500
=
548\,\mathrm{kB}
\end{equation}

of Flash. These footprint values are used for deterministic
proof-of-concept screening and are not reported as measurements obtained
from a deployed physical MCU.

Runtime measurements are obtained from a batch of 16 test images. Two
untimed warm-up executions are performed, followed by five repeated
measurements. The runtime proxy for each CAM-family configuration contains
the shared model forward-pass time, the base-CAM construction time, and
the configuration-specific post-processing time.

Within each stakeholder profile, runtime is min--max normalized over
configurations with available timing measurements, whereas the SRAM proxy
is normalized over configurations with available analytical SRAM metadata.
The relative deployment-cost proxy assigns weights of $w_T=0.85$ and
$w_M=0.15$ to runtime and SRAM, respectively. Configurations with incomplete
objective records are excluded before Pareto filtering. The resulting
quantity is a relative proxy rather than physical energy consumption or
MCU-measured latency.

\section{Results}

The predictive performance of the classifier was first evaluated on the
held-out test set. On the complete test set of 2,004 images, the
MobileNetV3-Small model achieved an accuracy of 66.77\% with a test loss
of 1.0641.

\subsection{LLM Proposals Across Stakeholder Profiles}

The top-five XAI methods proposed by GPT-4.1 mini and Gemini 2.0 Flash
were compared across the Clinical, TinyML, and Balanced stakeholder
profiles. Table~\ref{tab:llm_profile_proposals} reports the ranked
proposals and their execution scope.

\begin{table}[!t]
\centering
\caption{Top-five XAI methods proposed by GPT-4.1 mini and Gemini 2.0
Flash across the stakeholder profiles. Methods are listed in rank order;
$\dagger$ denotes host-only execution.}
\label{tab:llm_profile_proposals}

\scriptsize
\setlength{\tabcolsep}{3pt}
\renewcommand{\arraystretch}{1.15}

\begin{tabularx}{\columnwidth}{
>{\raggedright\arraybackslash}p{1.05cm}
>{\raggedright\arraybackslash}p{1.05cm}
>{\raggedright\arraybackslash}X
}
\toprule
\textbf{Profile}
& \textbf{Model}
& \textbf{Ranked proposals} \\
\midrule

Clinical
& GPT-4.1 mini
& 1) LR-CAM; 2) TopK-CAM; 3) Binary-CAM;
4) CAM; 5) GradCAM$^{\dagger}$ \\

& Gemini 2.0 Flash
& 1) LR-CAM; 2) CAM; 3) GradCAM$^{\dagger}$;
4) GradCAM++$^{\dagger}$; 5) Score-CAM$^{\dagger}$ \\

\midrule

TinyML
& GPT-4.1 mini
& 1) CAM; 2) LR-CAM; 3) Binary-CAM;
4) TopK-CAM; 5) Micro-CAM \\

& Gemini 2.0 Flash
& 1) CAM; 2) LR-CAM; 3) Binary-CAM;
4) TopK-CAM; 5) Micro-CAM \\

\midrule

Balanced
& GPT-4.1 mini
& 1) LR-CAM; 2) TopK-CAM; 3) Binary-CAM;
4) CAM; 5) GradCAM++$^{\dagger}$ \\

& Gemini 2.0 Flash
& 1) LR-CAM; 2) CAM; 3) TopK-CAM;
4) Micro-CAM; 5) GradCAM$^{\dagger}$ \\

\bottomrule
\end{tabularx}
\end{table}

\begin{itemize}

    \item \textbf{TinyML:}
    GPT-4.1 mini and Gemini 2.0 Flash proposed the same five methods in
    the same rank order: CAM, LR-CAM, Binary-CAM, TopK-CAM, and
    Micro-CAM. All five methods were classified in the method knowledge
    base as single-pass, on-device methods and passed the deterministic
    MCU-feasibility stage.

    \item \textbf{Clinical:}
    Both models ranked LR-CAM first. GPT-4.1 mini proposed four on-device
    methods and one host-only method, whereas Gemini 2.0 Flash proposed
    two on-device methods and three host-only methods. After deterministic
    MCU-feasibility filtering, LR-CAM, TopK-CAM, Binary-CAM, and CAM
    remained from the GPT-4.1 mini proposals, while LR-CAM and CAM
    remained from the Gemini proposals.

    \item \textbf{Balanced:}
    Both models proposed LR-CAM, CAM, and TopK-CAM. GPT-4.1 mini also
    proposed Binary-CAM and the host-only GradCAM++, whereas Gemini 2.0
    Flash also proposed Micro-CAM and the host-only GradCAM. The principal
    difference between the on-device proposal sets was therefore
    Binary-CAM in the GPT-4.1 mini shortlist and Micro-CAM in the Gemini
    shortlist. Four methods from each model passed deterministic
    MCU-feasibility filtering.

\end{itemize}

CAM-family methods were the most frequently proposed methods across the
three profiles in the observed runs. The two models produced identical
rankings for the TinyML profile but different method sets and rankings
for the Clinical and Balanced profiles.

\begin{figure*}[t]
    \centering
    \includegraphics[width=0.8\linewidth]
    {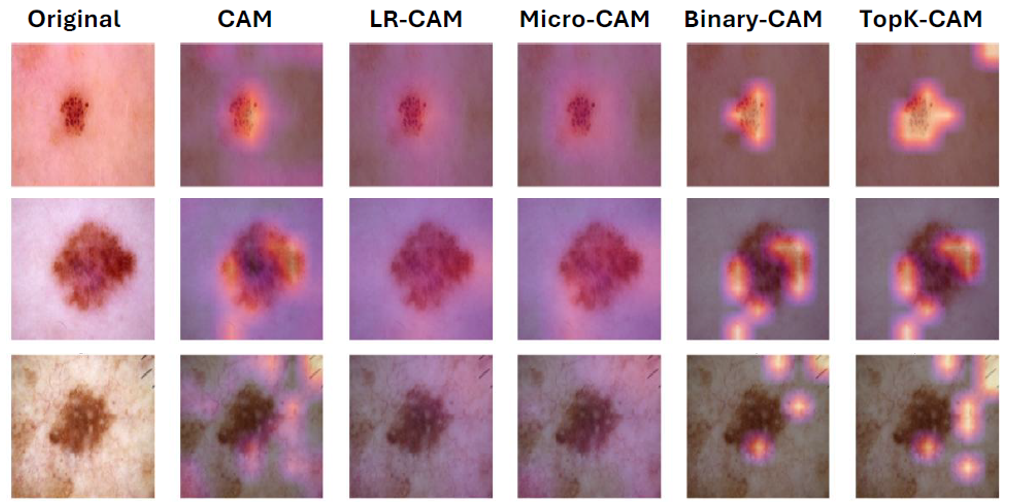}
    \caption{Representative outputs for the five XAI methods jointly
    proposed by GPT-4.1 mini and Gemini 2.0 Flash under the TinyML
    profile. The leftmost image is the original input, followed by CAM,
    LR-CAM, Micro-CAM, Binary-CAM, and TopK-CAM explanations.}
    \label{fig:gpt_tinyml}
\end{figure*}

Representative qualitative outputs for the five methods jointly proposed
by both models under the TinyML profile are shown in
Fig.~\ref{fig:gpt_tinyml}. These examples illustrate the visual effects of
the different CAM transformations; they are not presented as a clinical
validation of the resulting explanations.

\subsection{Runtime and Method-Level Overhead Analysis}

Runtime was measured in the experimental execution environment using a batch
of 16 test images. Two untimed warm-up iterations were performed before each
measurement. Each timing operation was then repeated five times, and the mean
runtime was recorded. All evaluated configurations used the shared-forward
execution model: a single model forward pass and base CAM computation were
shared with the configuration-specific post-processing operation.
Table~\ref{tab:latency_breakdown} reports the mean total shared-forward runtime
for representative configurations and the corresponding increase relative
to CAM within the same stakeholder profile.

\begin{table}[H]
\centering
\small
\caption{Mean shared-forward runtime for representative configurations.
Measurements were obtained for a batch of 16 images after two untimed
warm-up iterations and were repeated five times. Incremental overhead was
calculated relative to CAM within the same profile. All values are reported
in milliseconds.}
\label{tab:latency_breakdown}

\begin{tabular}{llcc}
\toprule
\textbf{Goal} &
\textbf{Method} &
\textbf{Total runtime} &
\textbf{Increase vs.\ CAM} \\
\midrule

\multirow{3}{*}{Clinical}
& CAM
& 45.38848
& 0.00000 \\

& \makecell[l]{Binary-CAM \\ ($\tau=0.80$)}
& 45.39458
& 0.00610 \\

& \makecell[l]{Binary-CAM \\ ($\tau=0.90$)}
& 45.39354
& 0.00506 \\

\midrule

\multirow{4}{*}{TinyML}
& CAM
& 43.20912
& 0.00000 \\

& \makecell[l]{Binary-CAM \\ ($\tau=0.60$)}
& 43.22048
& 0.01136 \\

& \makecell[l]{Binary-CAM \\ ($\tau=0.50$)}
& 43.22030
& 0.01118 \\

& \makecell[l]{Micro-CAM \\ ($s=5$)}
& 43.41238
& 0.20326 \\

\midrule

\multirow{4}{*}{Balanced}
& CAM
& 43.96584
& 0.00000 \\

& \makecell[l]{Binary-CAM \\ ($\tau=0.60$)}
& 43.97222
& 0.00638 \\

& \makecell[l]{Binary-CAM \\ ($\tau=0.50$)}
& 43.97226
& 0.00642 \\

& \makecell[l]{TopK-CAM \\ ($k=0.10$)}
& 44.09098
& 0.12514 \\

\bottomrule
\end{tabular}
\end{table}

The CAM baseline runtime differed slightly across the three profile-specific
timing runs. Comparisons were therefore made only relative to the CAM
measurement obtained within the same run. Among the representative
configurations, the largest measured increase was 0.20326~ms for Micro-CAM
with $s=5$ under the TinyML profile, followed by 0.12514~ms for TopK-CAM with
$k=0.10$ under the Balanced profile. The reported Binary-CAM configurations
added between 0.00506 and 0.01136~ms relative to their corresponding CAM
baselines. These batch-level measurements show that the shared model forward
pass and base CAM computation accounted for most of the total measured
runtime in the experimental environment. They should not be interpreted as
direct latency measurements on the target Cortex-M7 hardware.

\begin{figure}[H]
\centering
\includegraphics[width=\linewidth]{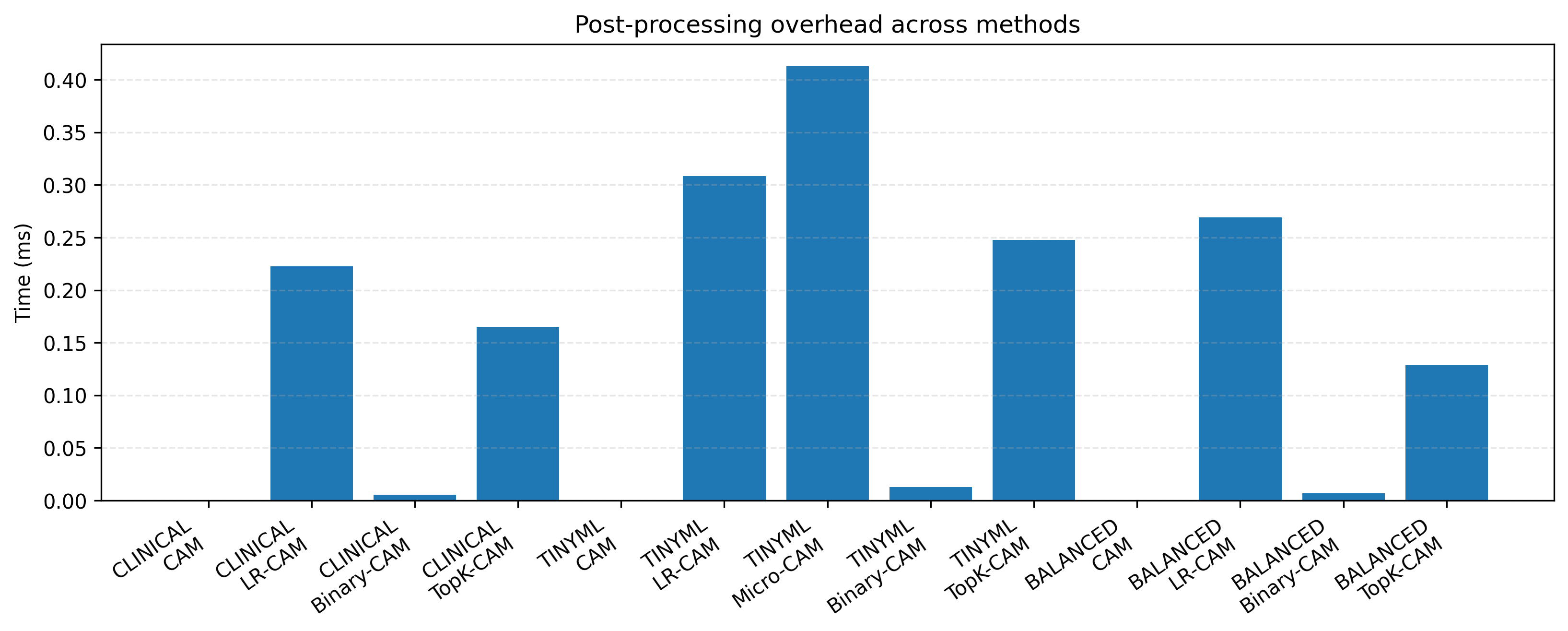}
\caption{Mean incremental post-processing overhead of each explanation
family relative to the CAM baseline, averaged across the evaluated parameter
settings. The reported values exclude neither the shared forward pass nor the
base CAM computation from the total-runtime measurements; the plotted
increment represents the configuration-specific increase relative to CAM.}
\label{fig:post_processing}
\end{figure}

Figure~\ref{fig:post_processing} complements the representative total-runtime
measurements by showing the mean incremental post-processing overhead of each
explanation family relative to CAM, averaged across its evaluated parameter
settings.

\subsection{Fidelity and Stability Evaluation}

The evaluation included 67 unique configurations under each of the three
deployment goals and used the complete held-out test set of 2,004 images.
For every configuration, the logit-based deletion AUC, insertion AUC, AOPC,
and composite fidelity values were identical across
\texttt{A2\_BALANCED}, \texttt{A2\_CLINICAL}, and
\texttt{A2\_TINYML}. This occurred because the same trained model, test set,
explanation configuration, and fidelity-evaluation procedure were used for
all three profiles. SSIM-based stability was evaluated in separate
profile-specific runs. Table~\ref{tab:fidelity_stability} therefore reports
the mean SSIM across the three profiles together with the corresponding
minimum and maximum values.

The highest composite fidelity score of 0.9397 was achieved by CAM, LR-CAM
with $d=1$, and Micro-CAM with sizes 7--10.
Table~\ref{tab:fidelity_stability} reports $s=10$ as a representative
configuration. These configurations produced a deletion AUC of 0.9745, an
insertion AUC of 1.9960, and an AOPC of 1.3302. However, the individual
fidelity metrics did not identify a single dominant configuration.
Micro-CAM with $s=6$ achieved the highest insertion AUC of 2.0021, whereas
TopK-CAM with $k=0.46$ achieved both the lowest deletion AUC of 0.9342 and
the highest AOPC of 1.3703. The latter nevertheless produced a lower
composite fidelity score and substantially lower SSIM-based stability.
These findings show that the fidelity metrics capture complementary aspects
of attribution behaviour and should therefore be interpreted jointly.

\begin{table}[H]
\centering
\caption{Representative fidelity and stability results across the Clinical,
TinyML, and Balanced profiles. Fidelity metrics were identical across
profiles, while SSIM is reported as the mean and range across the three
profile-specific runs. Higher values are preferable except for deletion AUC.
Bold values indicate the best observed result for each metric.}
\label{tab:fidelity_stability}

\scriptsize
\setlength{\tabcolsep}{1.6pt}
\renewcommand{\arraystretch}{1.08}

\begin{tabularx}{\columnwidth}{
>{\raggedright\arraybackslash}p{0.235\columnwidth}
>{\centering\arraybackslash}p{0.105\columnwidth}
>{\centering\arraybackslash}p{0.105\columnwidth}
>{\centering\arraybackslash}p{0.095\columnwidth}
>{\centering\arraybackslash}p{0.105\columnwidth}
>{\centering\arraybackslash}X
}
\toprule
\textbf{Configuration}
& \textbf{Del. $\downarrow$}
& \textbf{Ins. $\uparrow$}
& \textbf{AOPC $\uparrow$}
& \textbf{Fid. $\uparrow$}
& \textbf{SSIM mean [range] $\uparrow$} \\
\midrule

CAM
& 0.9745
& 1.9960
& 1.3302
& \textbf{0.9397}
& 0.8205 [0.8196, 0.8216] \\

LR-CAM ($d=1$)
& 0.9745
& 1.9960
& 1.3302
& \textbf{0.9397}
& 0.8193 [0.8166, 0.8211] \\

Micro-CAM ($s=10$)
& 0.9745
& 1.9960
& 1.3302
& \textbf{0.9397}
& 0.8201 [0.8174, 0.8236] \\

Micro-CAM ($s=6$)
& 1.1000
& \textbf{2.0021}
& 1.2046
& 0.7665
& 0.8663 [0.8640, 0.8698] \\

TopK-CAM ($k=0.46$)
& \textbf{0.9342}
& 1.8683
& \textbf{1.3703}
& 0.9209
& 0.6566 [0.6533, 0.6614] \\

Binary-CAM ($\tau=0.50$)
& 1.0201
& 1.8406
& 1.2850
& 0.7840
& 0.8252 [0.8200, 0.8296] \\

Micro-CAM ($s=4$)
& 1.2111
& 1.9926
& 1.0934
& 0.6042
& 0.8780 [0.8751, 0.8797] \\

Binary-CAM ($\tau=0.90$)
& 1.1506
& 1.6877
& 1.1549
& 0.5100
& 0.9271 [0.9252, 0.9291] \\

TopK-CAM ($k=0.04$)
& 1.1803
& 1.6527
& 1.1255
& 0.4477
& 0.9434 [0.9428, 0.9439] \\

Micro-CAM ($s=1$)
& 1.3724
& 1.4381
& 0.9333
& 0.0502
& \textbf{1.0000 [1.0000, 1.0000]} \\

\bottomrule
\end{tabularx}
\end{table}

The tested configurations also show that stronger spatial compression or
sparsification can increase SSIM-based stability while reducing fidelity.
Reducing the Micro-CAM resolution from $s=10$ to $s=1$ decreased composite
fidelity from 0.9397 to 0.0502 while increasing mean SSIM from 0.8201 to
1.0000. Similarly, reducing the TopK-CAM retained ratio from $k=0.46$ to
$k=0.04$ decreased fidelity from 0.9209 to 0.4477 while increasing mean
SSIM from 0.6566 to 0.9434. Increasing the Binary-CAM threshold from
$\tau=0.50$ to $\tau=0.90$ produced the same pattern, reducing fidelity
from 0.7840 to 0.5100 while increasing mean SSIM from 0.8252 to 0.9271.
A high SSIM value should therefore not be interpreted independently as
evidence of a high-quality explanation. Strongly compressed or nearly
invariant attribution maps may appear highly stable while retaining little
informative spatial variation. Fidelity and stability must be considered
jointly to distinguish informative and stable explanations from stable but
low-fidelity attribution maps.

\subsection{Pareto-Optimal Trade-Off Analysis}

The joint trade-off among fidelity, stability, and deployment cost was
examined using three-objective Pareto dominance. Composite fidelity and
SSIM-based stability were maximized, whereas the deployment-cost proxy was
minimized. Although fidelity and stability were available for all 67
configurations under each profile, the Pareto analysis required complete
fidelity, stability, and deployment-cost records. After this completeness
requirement was applied, 55 configurations were eligible under TinyML and
45 configurations were eligible under each of the Clinical and Balanced
profiles.

\begin{figure}[H]
\centering

\begin{subfigure}[t]{\linewidth}
\centering
\includegraphics[width=0.9\linewidth]
{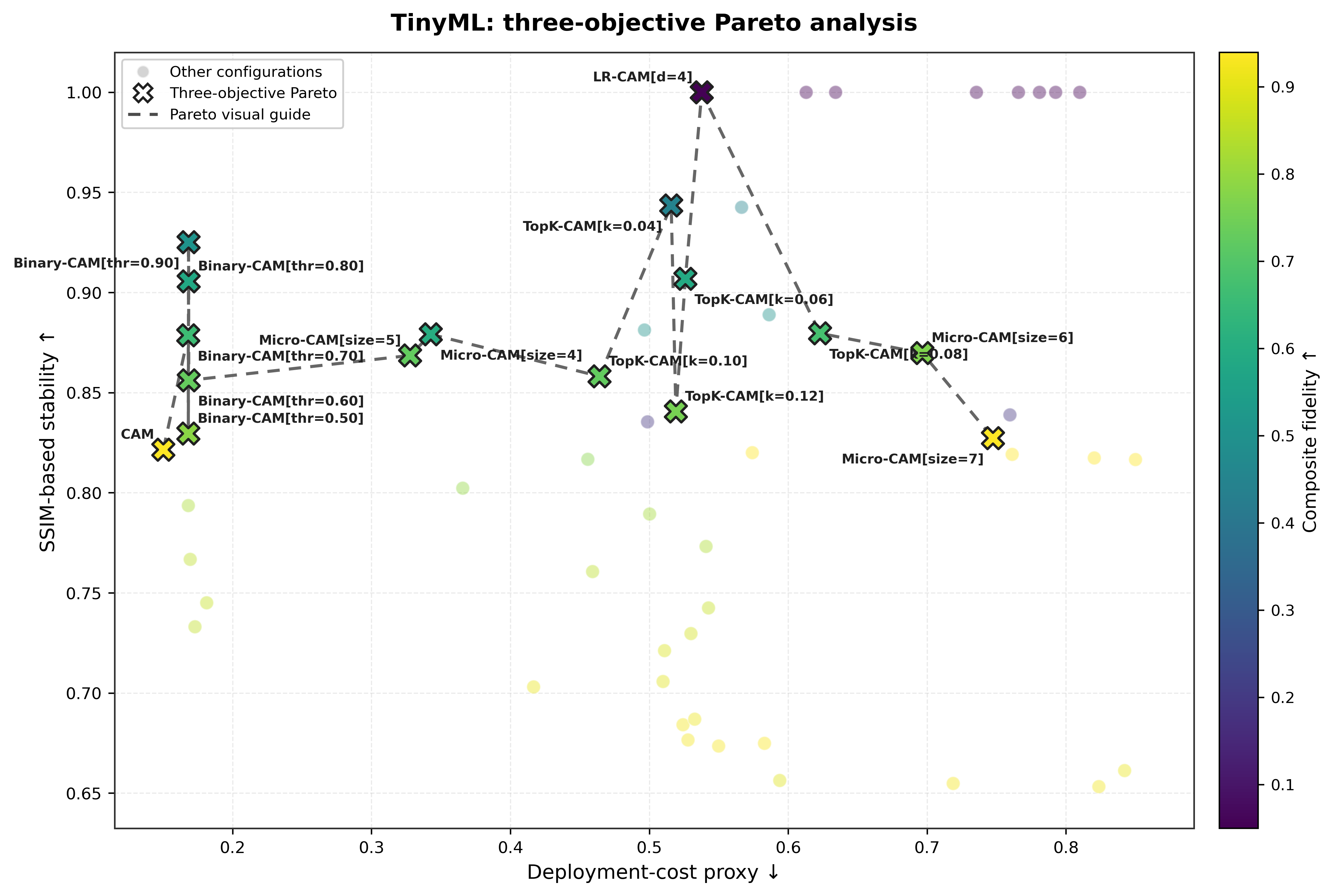}
\caption{TinyML}
\end{subfigure}

\begin{subfigure}[t]{\linewidth}
\centering
\includegraphics[width=0.9\linewidth]
{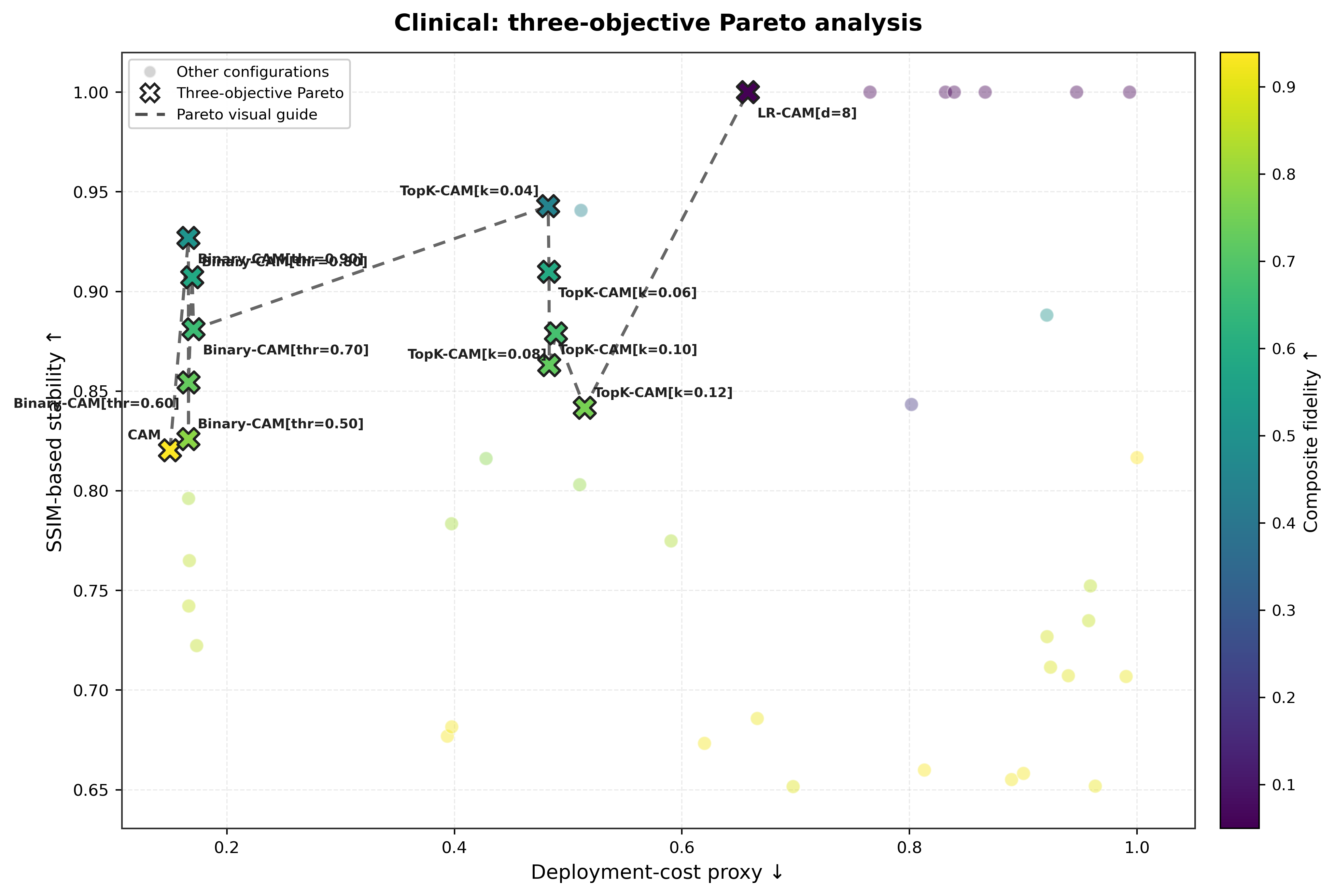}
\caption{Clinical}
\end{subfigure}

\begin{subfigure}[t]{\linewidth}
\centering
\includegraphics[width=0.9\linewidth]
{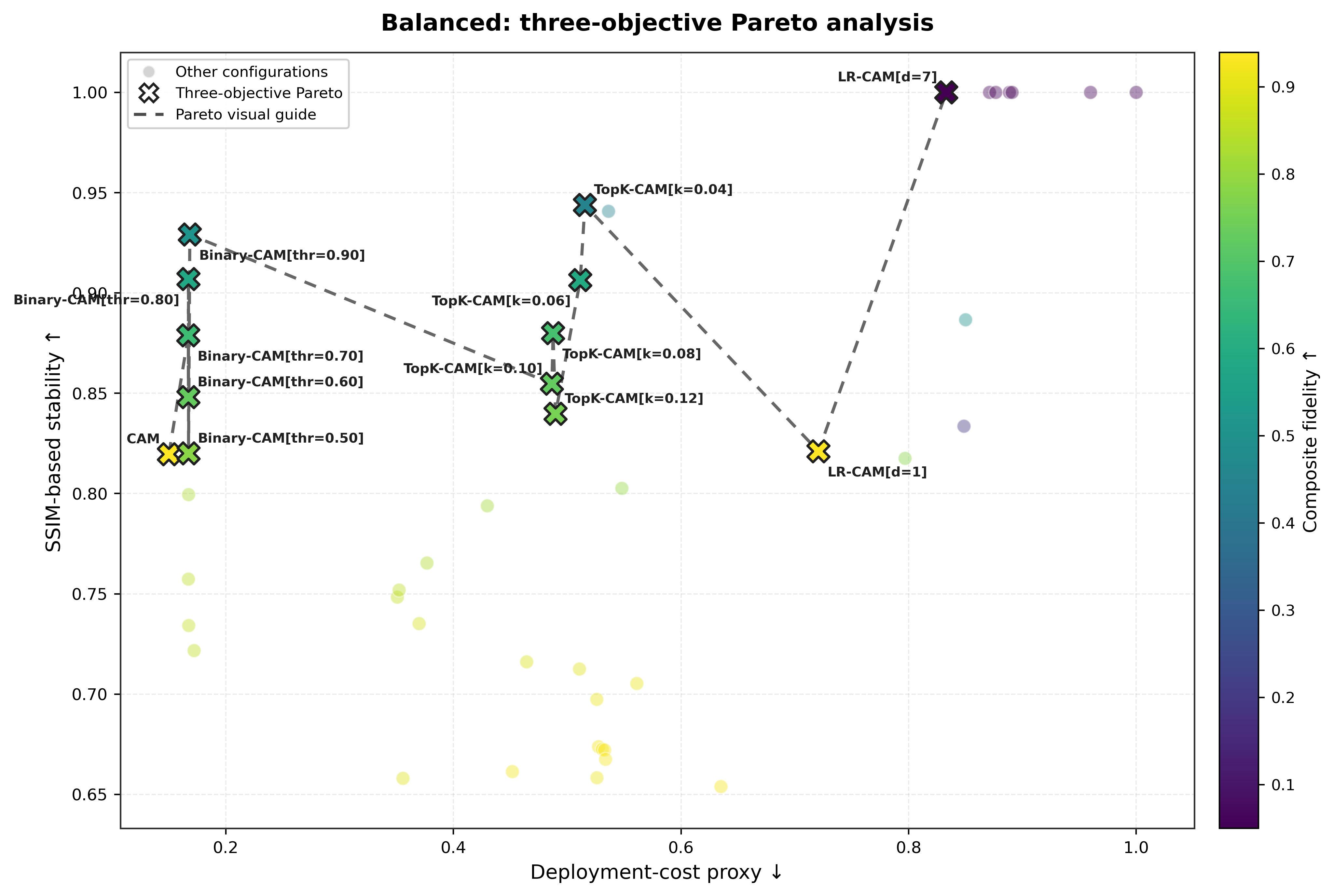}
\caption{Balanced}
\end{subfigure}

\caption{Two-dimensional visualizations of the three-objective Pareto
analysis for the (a) TinyML, (b) Clinical, and (c) Balanced profiles.
The horizontal and vertical axes represent the deployment-cost proxy and
SSIM-based stability, respectively, while point colour represents composite
fidelity. Cross markers identify configurations that are non-dominated when
deployment cost, fidelity, and stability are considered jointly. The dashed
lines connect these three-objective Pareto configurations in ascending
deployment-cost order and are included only as visual guides; they do not
represent a separately calculated two-objective Pareto frontier.}
\label{fig:pareto_all}
\end{figure}

Three-objective Pareto filtering retained 16 of the 55 eligible
configurations for TinyML, 12 of the 45 configurations for Clinical, and
13 of the 45 configurations for Balanced. In addition to this
three-objective analysis, Fig.~\ref{fig:pareto_all} presents a
two-dimensional visualization of deployment cost against stability. The
horizontal axis represents the deployment-cost proxy, the vertical axis
represents SSIM-based stability, and point colour encodes composite
fidelity. The highlighted cross markers are determined using all three
objectives rather than only the two displayed axes.
CAM appeared in the Pareto-optimal set under all three profiles at the
lowest deployment-cost value of 0.150. It combined the highest composite
fidelity score of 0.9397 with an SSIM-based stability value of approximately
0.82, making it a low-cost, high-fidelity reference configuration.

Binary-CAM variants occupied the low-deployment-cost region of each
Pareto-optimal set, with cost values of approximately 0.166--0.171.
Across the Pareto-optimal Binary-CAM configurations, increasing the binary
threshold generally increased stability from approximately 0.82 to 0.93
while reducing fidelity from 0.7840 to 0.5100. These configurations
therefore represented low-cost alternatives with different
fidelity--stability balances.
TopK-CAM variants occupied an intermediate deployment-cost region. Across
the three profiles, the Pareto-optimal TopK-CAM configurations had
deployment-cost values ranging from approximately 0.46 to 0.62 and
stability values between approximately 0.84 and 0.94. Lower retained ratios
generally produced higher stability but lower fidelity. For example,
TopK-CAM with $k=0.04$ achieved a stability value close to 0.94 but a
fidelity score of 0.4477, whereas TopK-CAM with $k=0.12$ retained a higher
fidelity score of 0.7578 with lower stability. This pattern illustrates the
trade-off introduced by increasingly sparse activation selection.

The TinyML Pareto-optimal set additionally included Micro-CAM configurations
with sizes 4--7. Their deployment-cost values ranged from approximately
0.33 to 0.75 and represented different compromises among spatial
compression, fidelity, and stability. Micro-CAM configurations had complete
timing and deployment-cost records only under the TinyML profile and were
therefore eligible for the Pareto analysis only in that profile. Their
absence from the Clinical and Balanced Pareto-optimal sets should not be
interpreted as a dominance-based rejection under those profiles.

Several LR-CAM configurations produced the maximum SSIM value of 1.000.
Among these configurations, the three-objective Pareto procedure retained
$d=4$ for TinyML, $d=8$ for Clinical, and $d=7$ for Balanced. Each of these
configurations had a composite fidelity score of only 0.0502. Their presence
in the Pareto-optimal sets should therefore not be interpreted as evidence
of superior explanation quality. Instead, they illustrate that an almost
invariant attribution map can maximize SSIM-based stability while retaining
little informative spatial variation. This finding reinforces the need to
consider fidelity, stability, and deployment cost jointly.

The Pareto-optimal sets shared a common low-cost structure based on CAM,
Binary-CAM, and TopK-CAM, but they were not identical. Micro-CAM appeared
only in the TinyML analysis because complete deployment-cost records were
available only for that profile, and the retained LR-CAM configuration
differed across the three profiles. No single configuration simultaneously
optimized all three objectives. The resulting non-dominated sets instead
provide multiple candidates for the subsequent stakeholder-specific
decision stage.

\subsection{Final Pareto-Constrained Configuration Selection}

Following the three-objective Pareto analysis, the non-dominated
configurations for each stakeholder profile were ranked using the
profile-specific goal score and supplied to GPT-4.1 mini for final
selection. The LLM was restricted to selecting exact configuration
identifiers from the corresponding Pareto-optimal set. It was permitted
to select one primary configuration for the Clinical profile and two
complementary primary configurations plus one fallback for the Balanced
and TinyML profiles.

The LLM output was subsequently subjected to deterministic validation.
For the Balanced and TinyML profiles, the complementary primary
configuration was required to retain a composite fidelity score of at
least 0.70, improve SSIM-based stability by at least 0.02 relative to the
high-fidelity anchor, and remain within 0.05 of the anchor's normalized
deployment-cost value. The fallback was required to be distinct from the
primary configurations, provide higher fidelity than the complementary
primary configuration, and remain within the same low-cost interval.
Every primary and fallback configuration was also required to belong to
the corresponding three-objective Pareto-optimal set.

GPT-4.1 mini produced a valid selection for all three profiles on the first
attempt. All selected configurations passed deterministic validation, and
no corrective retry was required. Table~\ref{tab:final_selection}
summarizes the resulting selections.

\begin{table}[H]
\centering
\caption{Final Pareto-constrained configurations selected by GPT-4.1 mini
for each stakeholder profile. Every primary and fallback configuration
belongs to the corresponding three-objective Pareto-optimal set.}
\label{tab:final_selection}

\small
\setlength{\tabcolsep}{4pt}
\renewcommand{\arraystretch}{1.10}

\begin{tabular}{lcc}
\toprule
\textbf{Profile}
& \textbf{Primary configurations}
& \textbf{Fallback} \\
\midrule

Clinical
& CAM
& -- \\

Balanced
& \makecell{CAM; Binary-CAM\\($\tau=0.60$)}
& \makecell{Binary-CAM\\($\tau=0.50$)} \\

TinyML
& \makecell{CAM; Binary-CAM\\($\tau=0.60$)}
& \makecell{Binary-CAM\\($\tau=0.50$)} \\

\bottomrule
\end{tabular}
\end{table}

Under the Clinical profile, GPT-4.1 mini selected CAM as the single final
configuration. CAM achieved the highest composite fidelity score of
0.9397, an SSIM-based stability value of 0.8203, and the minimum
deployment-cost value of 0.150. Although other Pareto-optimal
configurations achieved higher stability, they did so with lower
fidelity. CAM was therefore selected as the fidelity-oriented, low-cost
configuration for the Clinical profile.

Under the Balanced profile, CAM was selected as the high-fidelity,
low-cost anchor. Binary-CAM with $\tau=0.60$ was selected as the
complementary primary configuration. It increased stability from 0.8196
for CAM to 0.8481 while retaining a composite fidelity score of 0.7312
and a deployment-cost value of approximately 0.1672. Binary-CAM with
$\tau=0.50$ was selected as the fallback because it provided a higher
fidelity score of 0.7840, a stability value of 0.8200, and a similar
deployment-cost value of approximately 0.1673.

The same configuration structure was selected under the TinyML profile.
CAM provided a composite fidelity score of 0.9397, a stability value of
0.8216, and the minimum deployment-cost value of 0.150. Binary-CAM with
$\tau=0.60$ was selected as the complementary primary configuration,
providing a higher stability value of 0.8562 while retaining a fidelity
score of 0.7312 and a deployment-cost value of approximately 0.1683.
Binary-CAM with $\tau=0.50$ was selected as the fallback, providing a
higher fidelity score of 0.7840, a stability value of 0.8296, and a
deployment-cost value of approximately 0.1680.

CAM was selected across all three profiles as the common high-fidelity,
low-cost anchor. Under the Balanced and TinyML profiles, Binary-CAM with
$\tau=0.60$ provided a complementary increase in stability while
satisfying the specified fidelity and deployment-cost constraints.
Binary-CAM with $\tau=0.50$ provided a nearby Pareto-valid fallback with
higher fidelity than the complementary primary configuration. The final
LLM selections therefore retained distinct operating options within the
low-deployment-cost region while remaining subject to explicit and
deterministically validated selection constraints.

\section{Discussion}

The results expose a clear limitation of stability-only evaluation.
Micro-CAM with $s=1$ achieved an SSIM of 1.000 but a composite fidelity of
only 0.0502, whereas CAM achieved the highest fidelity of 0.9397 with an
SSIM of approximately 0.82. Binary-CAM and TopK-CAM showed the same
pattern: stronger sparsification increased stability while reducing
fidelity. SSIM must therefore be interpreted together with fidelity to
avoid favoring nearly invariant but weakly informative attribution maps.

The deployment measurements also distinguish spatial compression from
runtime efficiency. Binary-CAM retained the same estimated additional SRAM
as CAM (27.5625\,kB) but introduced only 0.00506--0.01136\,ms of additional
batch runtime in the reported configurations. Micro-CAM reduced the SRAM
estimate to 9.0\,kB but produced a larger increase of 0.20326\,ms for
$s=5$. Thus, reduced representation size did not necessarily correspond to
lower measured post-processing time.

Three-objective Pareto filtering retained 12 of 45 eligible Clinical
configurations, 13 of 45 Balanced configurations, and 16 of 55 TinyML
configurations. CAM was Pareto-optimal under all profiles, combining the
minimum deployment-cost proxy of 0.150 with the highest fidelity of 0.9397.
LR-CAM configurations with SSIM equal to 1.000 also remained Pareto-optimal,
but their fidelity of 0.0502 confirms that Pareto membership alone does not
indicate explanation quality. Micro-CAM appeared only in the TinyML analysis
because complete timing and cost records for this family were available only
for that profile.

GPT-4.1 mini selected CAM as the Clinical configuration and selected CAM
with Binary-CAM at $\tau=0.60$ for Balanced and TinyML, with
Binary-CAM at $\tau=0.50$ as the fallback. All selected configurations were Pareto-valid and passed deterministic
validation on the first attempt; no corrective retry was required. This
preserves a clear division between LLM-guided selection and deterministic
enforcement of Pareto membership, numerical constraints, and output
validity.

The runtime and deployment-cost results remain comparative rather than
hardware-level measurements. Timing was obtained in the experimental
execution environment, not on the target Cortex-M7, and the deployment-cost
objective used predefined SRAM estimates rather than measured energy.
Moreover, the heatmaps were not assessed using lesion masks or clinician
evaluation. The conclusions are therefore limited to the implemented
fidelity metrics, SSIM perturbations, timing measurements, and
deployment-cost proxy.
Future work should repeat the timing analysis on physical MCU hardware,
obtain parameter-specific memory and energy measurements, evaluate the
heatmaps using lesion annotations and dermatologists, and examine
sensitivity across repeated LLM runs and additional model architectures.

\section{Conclusion}

This study presented a human-centered framework for selecting XAI
configurations under TinyML deployment constraints. Across 67 evaluated
configurations, CAM provided the highest composite fidelity and the minimum
deployment-cost proxy, while stronger compression and sparsification often
increased SSIM stability at the expense of fidelity. Three-objective Pareto
filtering retained 12 Clinical, 13 Balanced, and 16 TinyML configurations
from their eligible sets.

LLMs generated profile-specific method proposals, and GPT-4.1 mini selected
final configurations only from the validated Pareto sets. Feasibility,
Pareto membership, numerical selection constraints, and output validity
remained deterministic. The results constitute a traceable proof of concept;
physical MCU measurements and clinical evaluation remain necessary before
deployment.

\section*{Acknowledgements}
This work was supported by UK Research and Innovation through the EPSRC National Edge AI Hub for Real Data [grant number EP/Y028813/1].

\bibliographystyle{IEEEtran}
\bibliography{main}

\begin{IEEEbiographynophoto}{Zeinab Dehghani}
Zeinab Dehghani is an MRes student in Computer Science at the University of Hull. Her research focuses on trustworthy and explainable AI, LLM-guided XAI design, and AI assurance for resource-constrained and safety-critical environments. She can be contacted at zeinab.dehghani068@gmail.com.
\end{IEEEbiographynophoto}

\begin{IEEEbiographynophoto}{Dhavalkumar Thakker} is a Professor of Artificial Intelligence (AI) and the Internet of Things (IoT) at the University of Hull, where he leads a group focused on Responsible Artificial Intelligence. His research emphasizes AI Explainability, AI Safety, and Fairness. With nearly two decades of experience, Dhavalkumar has been at the forefront of innovative solutions through funded projects. His interdisciplinary research spans Generative AI and the applications of Edge computing alongside IoT technologies.  He has a track record in leveraging AI for Social Good, notably in Smart Cities, Digital Health, and the Circular Economy. Contact him at D.Thakker@hull.ac.uk
\end{IEEEbiographynophoto}

\begin{IEEEbiographynophoto}{Koorosh Aslansefat} is an assistant professor of computer science at the University of Hull, HU6 7RX Hull, U.K., affiliated with the Dependable Intelligent System Group. His research interests span artificial intelligence safety, Markov modeling, and real-time dependability analysis. Aslansefat received his PhD in computer science from the University of Hull. He is a Member of IEEE. Contact him at K.Aslansefat@hull.ac.uk
\end{IEEEbiographynophoto}

\begin{IEEEbiographynophoto}{Kuniko Paxton} is a Postdoctoral Edge AI Researcher at the University of Hull and a member of the National Edge AI Hub, focusing on the development of efficient and responsible machine learning methods for resource-constrained and decentralized environments. Contact her at k.paxton@hull.ac.uk
\end{IEEEbiographynophoto}

\begin{IEEEbiographynophoto}{Bhupesh Kumar Mishra} is a Lecturer in AI and Data Science at the University of Hull. His research interests include Explainable
AI, Edge Computing, and the Internet of Things. Dr. Mishra received his PhD in Relied Item Optimization in Disaster Scenarios using computational algorithms from the University of the West of Scotland, UK. Contact him at Bhupesh.Mishra@hull.ac.uk
\end{IEEEbiographynophoto}

\begin{IEEEbiographynophoto}{Baseer Ahmad} is a Lecturer at
the University of Hull. His research interests include intelligent predictive
maintenance, the Internet of Things, and embedded electronic systems. Dr. Ahmad received his PhD in Predictive Maintenance from the University of the West of Scotland, UK. Contact him at Baseer.Ahmad@hull.ac.uk.
\end{IEEEbiographynophoto}

\begin{IEEEbiographynophoto}{Rameez Raja Kureshi} is a Lecturer and Program Director at the University of Hull, UK. His research interests include Edge AI, fairness, Cyber Security, and the Internet of Things. Dr. Kureshi received
his PhD in Artificial Intelligence and IoT from the University of Bradford, UK. He is a Senior Fellow at Advance HE. Contact him at R.Kureshi@hull.ac.uk.
\end{IEEEbiographynophoto}

\end{document}